\documentclass[journal,twoside,web]{ieeecolor}
\usepackage{jsen}
\usepackage{cite}
\usepackage{amsmath,amssymb,amsfonts}
\usepackage{algorithmic}
\usepackage{graphicx}
\usepackage{textcomp}
\usepackage{wrapfig}

\def\BibTeX{{\rm B\kern-.05em{\sc i\kern-.025em b}\kern-.08em
    T\kern-.1667em\lower.7ex\hbox{E}\kern-.125emX}}
\definecolor{abstractbg}{rgb}{0.89804,0.94510,0.83137}
\title{High-Precision UWB Sensor-Based Real-Time Locating System for Rodent Behavioral Studies}
\begin{document}
\author{
    Reza Sayfoori, 
    Mao-Hsiang Huang, 
    Amir Naderi, 
    Mehwish Bhatti, 
    Ron D. Frostig, 
    Hung Cao, \IEEEmembership{Senior Member,~IEEE}%
    \thanks{Corresponding author: Hung Cao. This work was supported in part by the National Institute of Neurological Disorders and Stroke (NINDS) under grant \#NS126526 (R.D.F.).}%
    \thanks{Reza Sayfoori, Mao-Hsiang Huang, Amir Naderi are with the Department of Electrical Engineering and Computer Science, University of California at Irvine, Irvine, CA 92697 USA (e-mail: rsayfoor@uci.edu).}%
    \thanks{Mehwish Bhatti is with the Department of Neurobiology and Behavior, University of California at Irvine, Irvine, CA 92697 USA (e-mail: mehwishb@uci.edu).}%
    \thanks{Ron D. Frostig is with the Department of Neurobiology and Behavior, and the Department of Biomedical Engineering, University of California at Irvine, Irvine, CA 92697 USA (e-mail: rfrostig@uci.edu).}%
    \thanks{Hung Cao is with the Department of Electrical Engineering and Computer Science, the Department of Biomedical Engineering, and the Department of Computer Science, University of California at Irvine, Irvine, CA 92697 USA (e-mail: hungcao@uci.edu).}
}

\IEEEtitleabstractindextext{%
\fcolorbox{abstractbg}{abstractbg}{%
\begin{minipage}{\textwidth}%
\begin{wrapfigure}[12]{r}{3in}%
\includegraphics[width=3in]{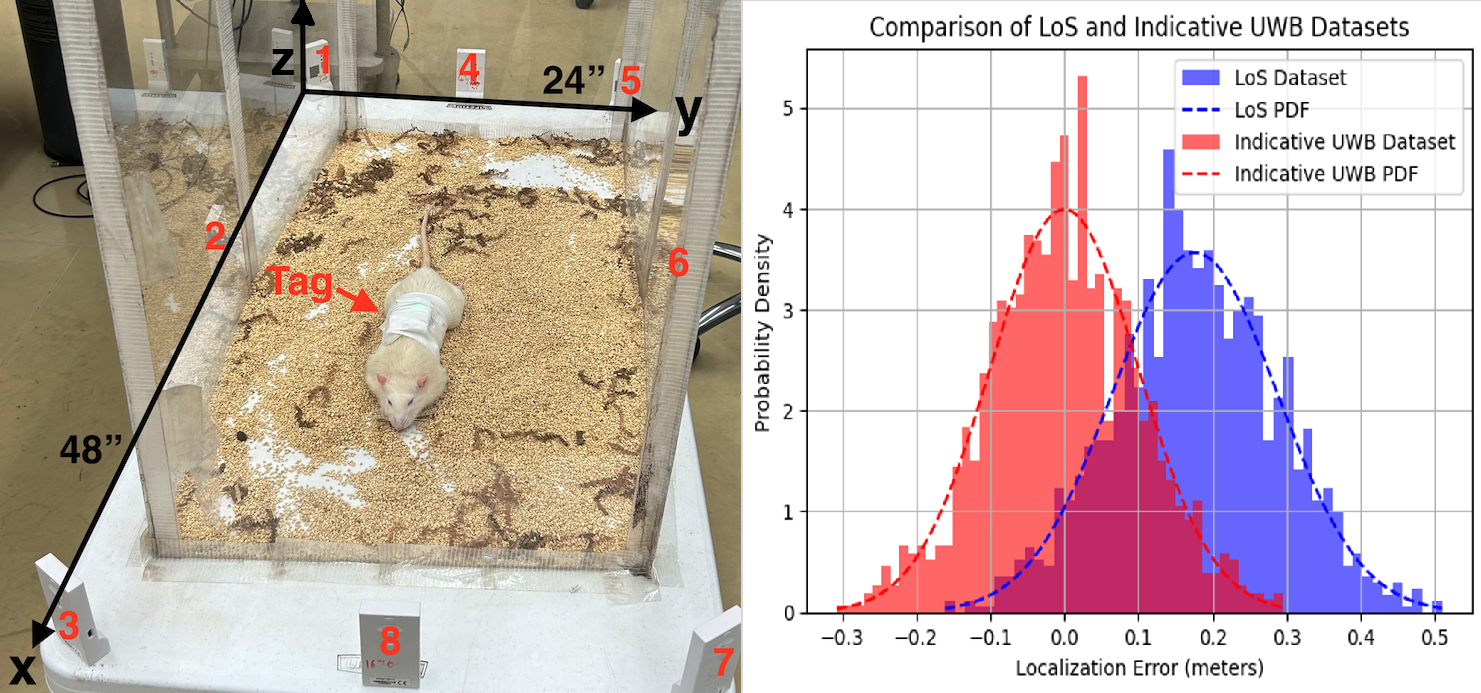}%
\end{wrapfigure}%
\begin{abstract}

Rodents have long been established as the premier model for behavioral studies, traditionally raised and maintained in conventional cage environments. However, these settings often limit rodents' ability to exhibit their full range of intrinsic behaviors and natural interactions. Precise tracking of animal movement is a critical component in behavioral research, but traditional methods, such as video tracking, present challenges, particularly with nocturnal species like rodents. This study introduces the application of ultra-wideband (UWB) sensor technology to develop a novel tracking system. The UWB DWM1001C sensor was integrated into a custom-made device worn by a single rat. A simplified habitat, measuring four-by-two feet, was used to evaluate system performance. The results show positioning accuracy errors of less than five millimeters for line-of-sight (LoS) and less than 50 millimeters for non-line-of-sight (NLoS) scenarios. This research provides a more accurate and reliable approach for animal localization, showcasing the potential of UWB sensor technology in enhancing precision in behavioral studies.

\end{abstract}

\begin{IEEEkeywords}
Ultra-Wideband (UWB) sensor technology, Line-of-Sight (LoS) sensor accuracy, Non-Line-of-Sight (NLoS) localization, Real-time UWB sensor-based animal tracking, High-precision UWB sensor systems.
\end{IEEEkeywords}
\end{minipage}}}

\maketitle
\section{Introduction}
\label{sec:introduction}
\IEEEPARstart{T}{he} aim is to replicate real-world living conditions for animals to explore behavior and neural processes \cite{b1}. In those settings, brain functions \cite{b0} and phenotypes are proven to be more intrinsic \cite{b2}, supporting various studies aiming to assess neurological and pharmaceutical effects in terms of movement, behavior, memory, anxiety, spatial orientation, and social interactions \cite {b3}. Among those, precise positioning and movement assessment are of utmost importance.
To gain accurate data, different methods for tracking animals using videos or radio frequency sensors have been deployed \cite{b4}. Video analysis for tracking animal movement has been widely employed since the early 1980s \cite{b5,b6}. The video tracking system integrates advanced sensor monitoring with high-resolution video, image processing, and real-time data analysis \cite{b7}\cite{b8}. This method requires specialized software \cite{b9} to capture detailed spatial information and analyze the collected data and sensor parameters \cite{b10}. However, since rodents are nocturnal and often remain in difficult-to-reach areas, this method is very limited \cite{b11}.
Radio frequency identification (RFID) sensor technology is an option for tracking rodents in experimental settings \cite{b12}. RFID sensor readers detect the signals transmitted by the tags and provide precise localization and interactions. However, RFID sensor limitations include the finite traveling signal range, interference of the signals with surroundings, limited data storage, power consumption in the active and passive tags, and most importantly, the line-of-sight (LoS) limitation \cite{b13}.

Collecting environmental sensor data such as sound, humidity, light, and temperature, and investigating how animals respond to their surroundings, is also necessary for behavioral studies \cite{b14}. Environmental sensor cues are monitored alongside neural activity and the behavior of the animal in real-time. Limitations may include interference from external electromagnetic fields in nearby devices, calibration, constant maintenance, cost, and data processing complexity, such as integrating and analyzing sensor data from multiple sources.

Multi-modal integration combines behavioral observation, environmental sensor data, and neural recording frameworks by synchronizing data from multiple sensor sources to correlate behavior and neural activities. This is another method to explore the mechanisms underlying animal behavior studies \cite{b15}. The challenges of this method may include synchronizing sensor data from multiple sources, such as environmental sensor data, neural recording, and behavioral observation, data processing complexity, interpretation complexity, correlation and causation, limited spatial and temporal resolution, and biological variability.

Wireless networks (WNs) and wireless sensor networks (WSNs) have gained a lot of attention in academic communities and industries to improve their application in this field \cite{b16}, especially for indoor positioning systems (IPS). Traditional tracking systems like inertial measurement units (IMUs) \cite{b17} and global positioning systems (GPS) \cite{b18} are designed and practiced outdoors, relying on satellite radio signals and dominating the navigation system in the air, seas, and any GPS devices with positioning accuracy of 4900 centimeters radius in open areas without obstacles \cite{b19}. Recent studies show the most popular use of wireless sensor technologies includes WIFI, Bluetooth, Zigbee, RFID, ultrasound, millimeter-wave (mm-wave), and ultra-wideband (UWB) sensors \cite{b20}. Wireless sensor communication technologies such as non-orthogonal multiple access (NOMA) \cite{b21}, mm-wave \cite{b22}, unmanned aerial vehicles (UAV) \cite{b23}, intelligent reflecting surfaces (IRS) \cite{b24}, tera-hertz (THz) communication systems, industrial internet of things (IIoT) \cite{b25}, and similar sensor technologies have dominated automation and process control systems in wireless communication.

Our study is focused on the UWB sensor technique for IPS, with its high channel capacity and extremely wide bandwidth signal that exceeds 500 MHz, low energy consumption that aligns with ITU-R \cite{b26}, short period of pulses to reduce multipath fading \cite{b27}, and its ability to be reprogrammed and used for different applications. Additionally, the time of arrival (ToA), the time difference of arrival (TDoA) signal, and its two-way time of arrival (TW-ToA) signal in UWB sensors make it a suitable technology for our delegate application \cite{b28}. We employed DecaWave DWM1001C sensor module kits integrated with an Android application for precise tracking of moving objects in dense environments, including tracking small animals like rats for behavioral studies. This novel approach to the tracking system and technique offers very precise live tracking techniques based on the DecaWave application, from setting up the tag and beacon sensors to Android applications and collecting live sensor data of moving objects. This approach reviews and improves the use of the time of flight (ToF) sensor signal \cite{b29}, LoS, and non-line-of-sight (NLoS) sensor signals \cite{b30} to achieve and mitigate the error of obstacles in the path of UWB sensors by using filtering applications to collect solid sensor data.

This work aims to improve IPS and wireless sensor applications by characterizing UWB sensor technology for accurate tracking. It examines the impact of UWB sensor signals in both LoS and NLoS scenarios and proposes methods to mitigate positioning errors. Additionally, the study focuses on enhancing signal accuracy and data collection for tracking lab animals, particularly in challenging environments. Efforts were also made to reduce tracking errors for moving objects in areas typically considered difficult to monitor.

\section{Materials and Methods}
\label{sec:Implementation}

\subsection{Hardware and System}
In this work, we use the DecaWave DWM1001C sensor by Qorvo, which integrates UWB sensor and Bluetooth antennas, RF circuitry, and the Nordic Semiconductor nRF52832 \cite{b31}. Segger Embedded Studio and J-Link are utilized for firmware updates. The Android application implements a real-time locating system (RTLS) for positioning and the PANS sensor communication protocol \cite{b32}. The antenna, with an omnidirectional radiation pattern and a peak gain of 2.5 dBi, operates in the 5.5 to 7.5 GHz range. The sensor kit can update tag locations at a frequency of 10 Hz. Each module is configurable as an anchor or tag via USB or Bluetooth using the DecaWave DRTLS application. Anchors serve as fixed sensor nodes, while tags are mobile sensor nodes. The UWB sensor technology, compliant with IEEE 802.15.4a and 802.15.4z standards, operates in the 3.1 to 10.6 GHz range with minimal interference, enabling accurate time measurement for localization within centimeter-level accuracy. The low power consumption of the sensor extends battery life, while its high data rate and short pulse response enhance data integrity and security. Communication between DWM1001C sensor nodes occurs at 150 Hz, managed by the PANS protocol \cite{b33}. This allows networks of up to 750 tags at 0.2 Hz or 15 tags at 10 Hz. The ranging algorithm includes time-based (ToA, TDoA, TW-ToA, PoA) and signal-based (RSSI, CSI) techniques. The DWM1001C sensor uses two-way ranging to calculate time of flight (ToF) for distance measurement through continuous data exchange between sensor nodes.

\vspace{-1mm}
\begin{equation}
R = \frac{(T_{rr} - T_{sp}) - T_{rsp} + (T_{rf} - T_{sr}) - T_{rsp}}{4}
\end{equation}

\vspace{-1mm}
where \( R \) represents the result of the calculation, \( T_{rr} \) is the time of the reflected signal at the receiver, \( T_{sp} \) is the time of the signal at the signal processor, \( T_{rsp} \) is the combined processing and reflection time, \( T_{rf} \) is the time at the receiver front, and \( T_{sr} \) is the time at the signal reflector.

\vspace{-2mm}
\begin{equation} 
 d = c \times t
\end{equation}

\vspace{-2mm}
where \(d \) is the distance, \( c \) is the speed of light (c), and \( t \) is the ToF.

The normalized distance (ND) and phase equation calculates the long distance within the network of each anchor to collect precise localization measurements. By integrating data analysis tools to improve and maximize the ability of any movement pattern of the object for data collection. 

\vspace{-1mm}
\begin{equation}
\text{ND} = \frac{\text{distance}}{\text{range}} = \text{phase} = \frac{T_{RT}}{T} = \frac{V_R - B}{2V_R - (A + B)}
\end{equation}

\vspace{-1mm}
where ND is defined as the ratio of the \textit{distance} to the \textit{range}. This is equivalent to the signal phase, expressed as the ratio of the round-trip time \(T_{RT}\) to the total time period \(T\). The final expression relates the velocities and distances involved, where \(V_R\) is the relative velocity, \(A\) is the initial distance, and \(B\) is the final distance, providing a simplified model for signal propagation dynamics.

\vspace{-2mm}
\subsection{Experimental Setup}
\begin{figure}[ht]
    \centering
    \includegraphics[width=1\linewidth]{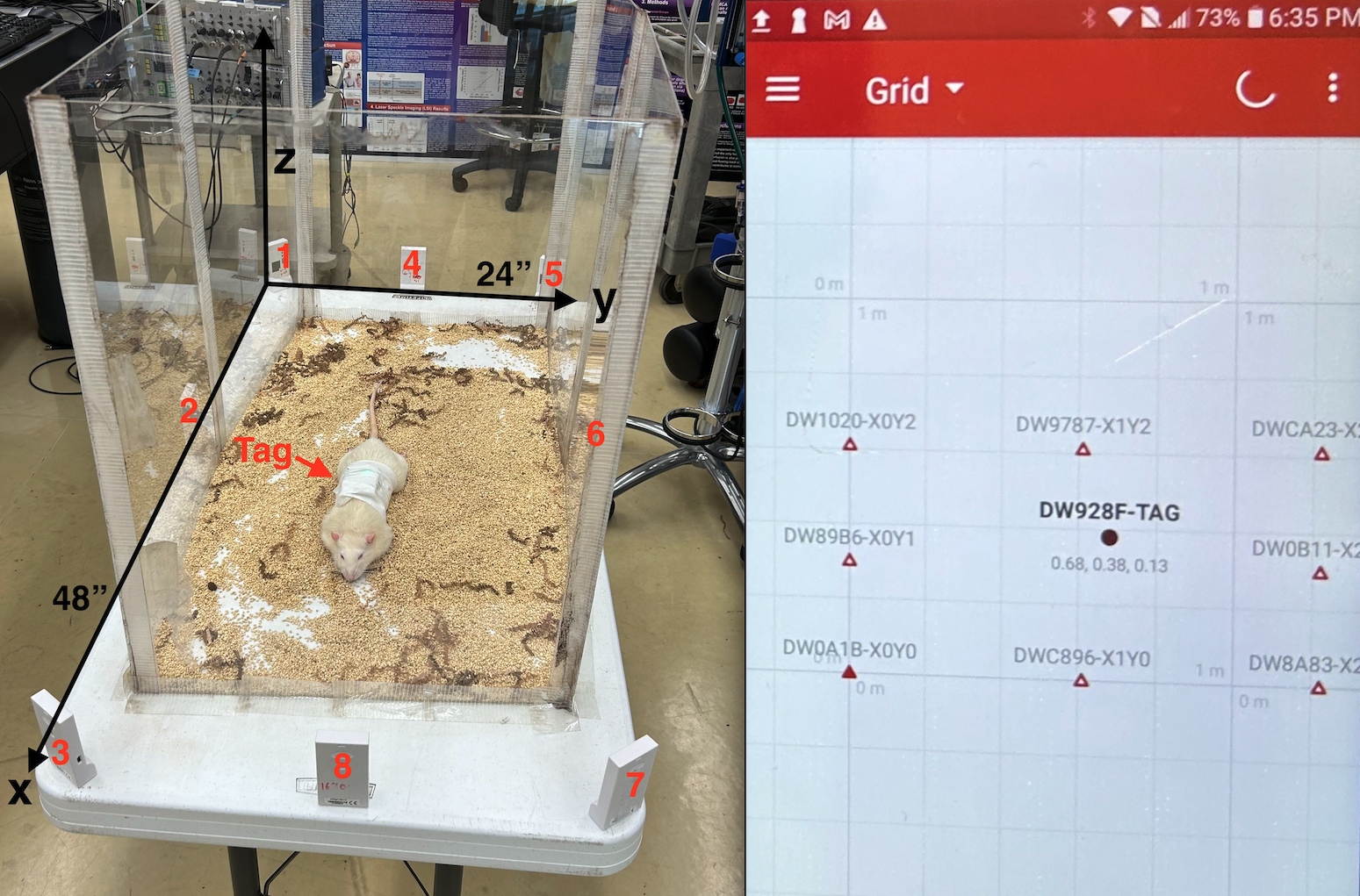}
    \caption{Figure illustrates the experimental setup for tracking a rat using UWB sensor technology. The enclosure measures 48 inches by 24 inches. A tagged rat (red arrow) moves within, with its position tracked by eight anchors (1-8) placed strategically around the perimeter. The right panel displays real-time coordinates of each anchor and the tag, facilitating precise location monitoring.}
    \label{fig:Rat}
\end{figure}

The test environment accrued in a cage built of epoxy glass with measurements of 24 inches by 48 inches in the biomedical lab environment, and by positioning eight sensor-equipped anchors in fixed places acting as sensor nodes, and a mobile sensor tag acting as the bridge attached to the rat to communicate between the UWB sensor network. The system is used to track any movement of the animal and to measure the accuracy or possible errors (Figure~\ref{fig:Rat}). The test was intended to evaluate the RTLS sensor regularity and its implementation, calculated as the ratio between successful sensor position determinations and the total trials.

\vspace{-2mm}

\vspace{-2mm}
\subsubsection{Network}
A network cluster of all anchors as sensor beacons and a sensor tag in passive mode as a sensor listener is created in the Android application (Figure~\ref{fig:Rat}). The sensor location network between PANS and DWM1001C sensor modules is based on time division multiple access (TDMA). At every instant of time, there will be only one sensor tag-anchor pair exchanging sensor packet communication messages to measure the sensor round-trip time or two-way ranging system. The node sensor is using a 100 msec long sensor super-frame, and the synchronization on the TDMA is capable of avoiding any sensor collision.

\subsubsection{Positioning Technique} 
The positioning technique relies on a controlled environment with the calculated and stationary location of multi-beacon sensors and moving tag sensors to transmit and receive location-based sensor data with its reference points. The calculation of objects in three-dimensional coordinates for (X, Y, Z) and to locate and track the object and its direction is based on sensor measurements. For this application, the ToA, TDoA, and phase measurement techniques have been employed to capitalize on UWB sensor accuracy and persistent tracking systems to minimize the interference and blockage of obstacles. The design of this study was to reduce the error generated by interference signals and obstacles and improve the accuracy of sensor signals for scientific study.

In the biomedical lab, using anesthesia by the medical professional, the sensor tag attached to the collected Rat and the real-time sensor visualization of Android provided the graphical representation for immediate error detection and examination of possible sensor glitches in the system. For this setup, the sensor calibration algorithm uses environmental factors to increase the accuracy of sensor positioning. The robust data security protocol and data sensitivity promised the integrity of the sensor dataset. In this experiment, the sensor channel capacity of bandwidth increased using the Shannon–Hartley theorem \cite{b34} capacity formula:

\vspace{-3mm}
\begin{equation}
C = B \cdot \log_2(1 + \text{SNR})
\label{eq:Shannon–Hartley_capacity}
\end{equation}

\vspace{-2mm}
where \(C \) means the channel capacity, \( B \) is the bandwidth of the channel, and \( SNR \) means the signal-to-noise ratio.

\vspace{-3mm}
\subsection{Methodology and Predictive Modeling}
By using the signal penetration depth equation and understanding the use of different frequencies we were able to model the signal strength in transmit and receive:

\vspace{-2mm}
\begin{equation}
    D = \frac{\lambda}{4\pi} \sqrt{\frac{Pt \cdot Gt \cdot Gr}{Pr}}
\end{equation}

\vspace{-1mm}
Here, \( D \) is the distance, \( \lambda \) the wavelength, \( P_t \) the transmitted power, \( G_t \) and \( G_r \) the antenna gains, and \( P_r \) the received power. This equation relates these parameters to determine the signal transmission range in free space. The link between bandwidth and spatial resolution was explored by associating the module's wide bandwidth with its spatial accuracy.

\vspace{-1mm}
\begin{equation}
    \ (\Delta x )= \frac{c}{2B}\
\end{equation}

\vspace{-1mm}
where \(\Delta x\) represents the resolution or accuracy of distance measurement, \(c\) is the speed of light, and \(B\) is the bandwidth of the UWB signal. The factor of \( \frac{1}{2} \) accounts for the round-trip travel time of the signal, as the distance is calculated based on the signal traveling to the object and back.

The module's \(ToF\) computation, essential in multi-layered environments, was analyzed by the following equation. 

\vspace{-2mm}
\begin{equation}
ToF = \frac{d}{c}
\end{equation}

\vspace{-1mm}
where \(d\) is the distance traveled by the signal, and \(c\) denotes the speed of light. 

The positioning accuracy of the sensor is crucial for objects like rats, was discussed using combining \(ToF\) measurement deviations (\( \sigma_{ToF} \)) and synchronization errors (\( \sigma_{sync} \)). 

\vspace{-2mm}
\begin{equation}
\sigma_{\text{pos}} = \sqrt{\sigma_{\text{ToF}}^2 + \sigma_{\text{sync}}^2}\
\end{equation}

\vspace{-1mm}
where \(\sigma_{\text{pos}}\) is the overall position uncertainty, \(\sigma_{\text{ToF}}\) denotes the uncertainty due to ToF measurements, and \(\sigma_{\text{sync}}\) accounts for the synchronization error.

\vspace{-4mm}
\subsection{Technical Insights of DecaWave DWM1001C Sensor}
Each anchor is placed in a fixed measured position, the mobile tag sensor is attached to the rat, and each coordinate is imported to the Android application by the origin of the X- and Y-axes. Also, by using the following equation we predict any path loss in wireless communication:

\vspace{-2mm}
\begin{equation}
A = 20 \log_{10}\left( \frac{4\pi d_f}{\lambda} \right) + n \cdot L \cdot F \
\end{equation}

\vspace{-1mm}
where \( A \) represents the attenuation in decibels (dB), \( d_f \) is the distance between the transmitter and receiver sensors, \( \lambda \) is the wavelength of the signal, \( n \) denotes the number of obstacles or environmental factors, \( L \) is the path loss coefficient, and \( F \) is the frequency-dependent loss factor. 

Considering distance \(d_{f}\), frequency \(F\), and wall loss \(L\). The strategic placement of anchors sensors for accurate positioning was examined through triangulation equations. 

\vspace{-3mm}
\begin{equation}
x = \frac{d_1^2 - d_2^2 + d_{12}^2}{2d_{12}}\
\end{equation}

\vspace{-3mm}
\begin{equation}
y = \frac{d_1^2 - d_3^2 + d_{13}^2 - x^2}{2d_{13}}\
\end{equation}

\vspace{-1mm}
where \(x\) and \(y\) represent the coordinates of the target, \(d_1\), \(d_2\), and \(d_3\) are the distances from the target sensor to three reference points, and \(d_{12}\), \(d_{13}\) are the distances between the reference points. The first equation calculates the \(x\)-coordinate using the distance differences, and the second equation derives the \(y\)-coordinate by adjusting for the already computed \(x\)-value.
\vspace{-5mm}
\subsection{Data analysis and comparative performance of RTLS}
The UWB sensor contains two signal classifications: LoS condition, in which the path of detection is completely clear without the interference of any obstacle between the sensor tag and sensor anchors. Additionally, NLoS condition, in which the sensor signal is reflected by an obstacle present between the sensor anchors and sensor tags in a very dense environment. By reducing the environmental interference and accounting for the sensor signal strength effect in a multipath area, and through dynamic sensor pulse repetition frequency adjustment, we enhanced the sensor signal quality in extremely unreachable locations. This enhancement includes the adoption of sensor antenna signals and sensor signal processing algorithms in NLoS conditions. The performance accuracy of sensor signal strength and system latency improved with measurable low error and robust sensor connectivity.

\vspace{-2mm}
\section{Results}
\label{sec:experiment}

\subsection{Comprehensive Analysis of UWB sensor in LoS Data Collection and Accuracy Assessment}

This study leverages the DWM1001C sensor modules to evaluate the UWB sensor LoS tracking alongside video-based tracking, revealing a high degree of accuracy within UWB sensor systems. Comparatively, minimal deviations were observed between UWB sensors in LoS coordinates and those recorded via video tracking, with an average difference of approximately 1.18 cm across selected data points, underscoring the potential of UWB sensors for high-precision sensor-based applications. The maximum deviation noted was 2.28 cm, which is within acceptable limits for most biomedical tracking applications in controlled sensor environments.

\begin{figure}[ht]
    \centering
    \includegraphics[width=1\linewidth]{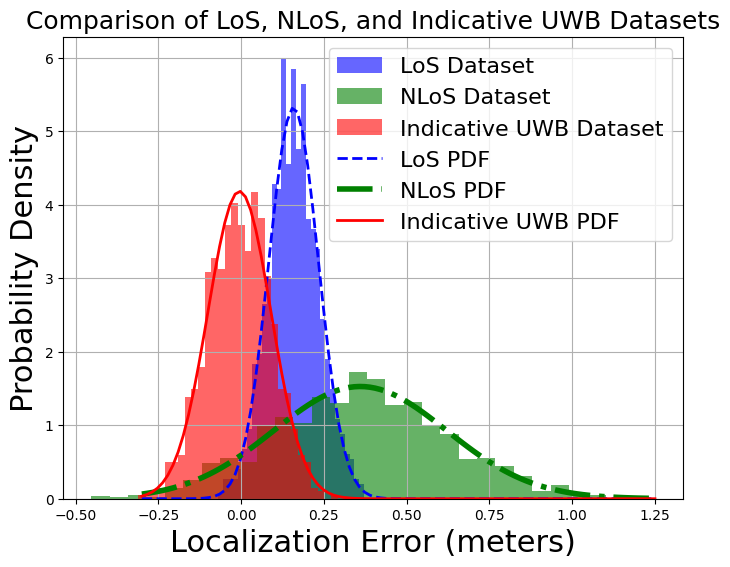}
    \caption{Comparison of localization error distributions between LoS, NLoS, and Indicative UWB datasets. The plot shows histograms of the localization errors for each dataset, with their corresponding Probability Density Functions (PDFs) fitted using normal distributions. The LoS dataset exhibits a narrower spread and lower mean error (mean = 0.162 m, $\sigma$ = 0.076 m), indicating higher accuracy. The NLoS dataset shows a wider spread and a higher mean error (mean = 0.356 m, $\sigma$ = 0.270 m), demonstrating the adverse effects of obstacles on tracking accuracy. The Indicative UWB dataset, included for comparison, shows a different error distribution with a mean of 0 m and $\sigma$ = 0.1 m.}
    \label{fig:LOS}
\end{figure}

In the context of LoS conditions, the dataset comprised 267 sensor entries, revealing a mean localization error of 0.162 m with a standard deviation (\(\sigma\)) of 0.076 m. This dataset is depicted in Figure~\ref{fig:LOS}, which employs a probability density function (PDF) \cite{b36} to model the variability of sensor error values.

\vspace{-2mm}

The close alignment between the UWB sensor in LoS and video sensor tracking outcomes not only validates the UWB sensor's efficacy but also supports its application as a reliable sensor tool for capturing fine-scale movements and spatial dynamics.

\begin{figure}[ht]
    \centering
    \includegraphics[width=0.82\linewidth]{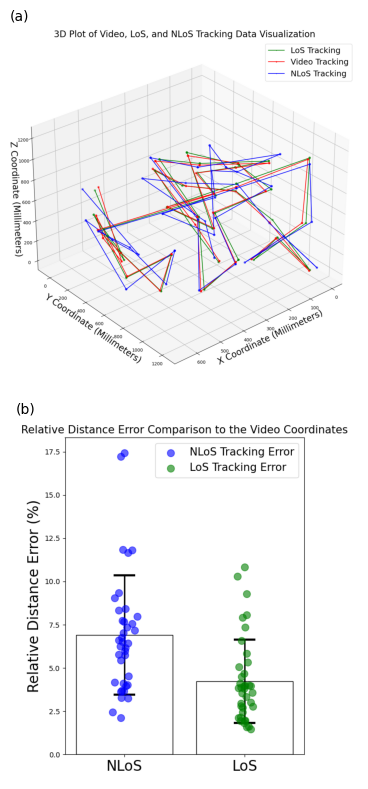}
    \caption{(a) 3D plot of animal movement tracked by video, UWB LoS, and UWB NLoS, with LoS closely aligned to video and NLoS showing deviations due to interference. (b) Bar plot comparing percentage distance error between UWB NLoS and LoS relative to video tracking, with bars showing mean error and error bars representing standard deviation.}
    \label{fig:3D}
\end{figure}

The plot in Figure~\ref{fig:3D} compares the performance of two sensor-based tracking methods, NLoS and LoS, against the ground truth provided by video tracking. The NLoS method shows a higher average sensor error and greater sensor variability compared to the LoS method. This suggests that the LoS tracking is more accurate and consistent, as expected given the clear line of sight between the transmitter and receiver sensors. In contrast, NLoS sensor tracking is is more accurate and consistent, as expected given the clear line of sight between the transmitter and receiver sensors. In contrast, NLoS sensor tracking is influenced by obstacles and signal reflections, resulting in a larger spread of sensor errors. The individual sensor data points highlight the variations within each method, where NLoS has a few high outliers that further inflate its average sensor error.
\vspace{-2mm}
\subsection{Data Collection under NLoS Conditions}
For the NLoS scenario, physical barriers such as furniture and surrounding lab instruments were introduced, causing signal reflections and obstruction. These NLoS conditions closely replicate real-world challenges, significantly impacting UWB sensor performance. The UWB sensor anchors were strategically placed around the test bench to simulate these obstruction patterns and assess sensor tracking accuracy under varying conditions.
For data collection, a combination of automated signal logging and manual video annotation was employed. UWB sensor receivers meticulously recorded the signal strength, quality, and time of arrival of each transmitted sensor signal. The video data, annotated frame-by-frame (fbf), served as the ground truth, allowing for an exact comparison between the observed and recorded sensor positions.

The examination of the data elucidated marked discrepancies in UWB sensor tracking accuracy under various conditions. Specifically, under NLoS conditions, the average deviation between UWB sensor measurements and video annotations was 35.8 millimeters, with deviations extending up to 61.2 millimeters in the most challenging scenarios. The standard deviation of these sensor measurements was 8.5 millimeters, indicating significant variability, likely due to multipath effects where UWB sensor transmitted signals reflect off multiple surfaces before reaching the sensor receiver.

In stark contrast, LoS conditions showed considerably less deviation, with an average of only 11.8 millimeters and a maximum deviation of 22.8 millimeters, coupled with a lower standard deviation of 3.5 millimeters. This pronounced disparity underscores the substantial impact of environmental obstructions on UWB sensor signal fidelity and sensor tracking accuracy, as detailed in Figure~\ref{fig:3D}.

Further analysis of the NLoS dataset, heavily influenced by physical barriers and consisting of 220 sensor entries, revealed a higher mean sensor error of 356 millimeters with a standard deviation of 270 millimeters. This highlights the adverse effects of environmental obstructions on UWB sensor tracking performance. The probability density function for the NLoS sensor errors, illustrating the distribution of error magnitudes, is defined as follows:

\vspace{-2mm}
\begin{equation}
P(x) = \frac{1}{\sigma\sqrt{2\pi}} e^{-\frac{(x-\mu)^2}{2\sigma^2}}
\end{equation}

\vspace{-2mm}
where \(\mu\) is the mean error and \(\sigma\) is the standard deviation, quantifying the spread of errors due to obstructions. This analysis clearly demonstrates how environmental conditions play a critical role in the accuracy of UWB tracking systems.

\vspace{-2mm}
\subsection{SNR Calculation for Performance Analysis}
To evaluate signal quality in noisy environments, the signal-to-noise ratio (SNR) \cite{b38} is used, defined as the ratio of signal power to noise power, expressed in decibels (dB). The SNR is calculated as:
\vspace{-4mm}
\[\text{SNR (dB)} = 10 \log_{10} \left(\frac{P_{\text{signal}}}{P_{\text{noise}}}\right)\]

\begin{figure}[ht]
    \centering
    \includegraphics[width=1\linewidth]{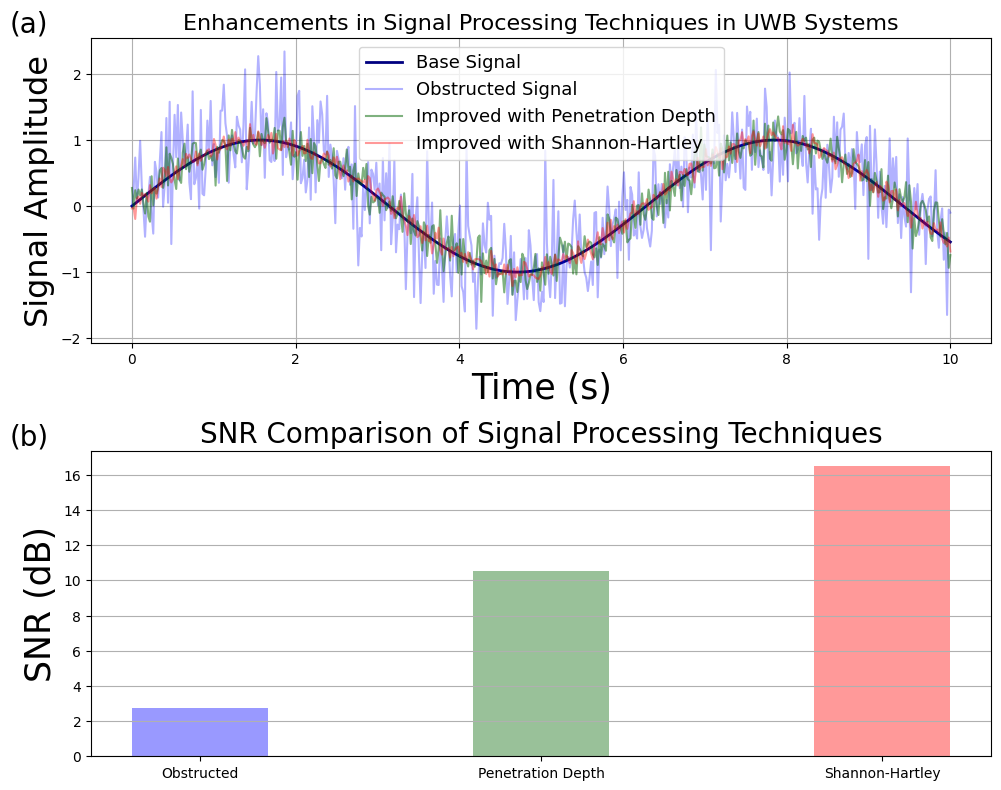}
    \caption{(a) Signal enhancement techniques for UWB sensor systems, comparing base, obstructed, and improved signals. (b) SNR comparison of the obstructed signal, improvement via penetration depth, and Shannon-Hartley theorem.}
    \label{fig:SNR}
\end{figure}

\vspace{-2mm}
where \( P_{\text{signal}} \) and \( P_{\text{noise}} \) represent the power (variance) of the signal and noise, respectively. In practical applications, such as UWB sensor systems, SNR is crucial for assessing system performance in the presence of noise. The SNR can be plotted to compare signal performance under various conditions, as shown in Figure \ref{fig:SNR}.
\vspace{-2mm}

\vspace{-3mm}
\subsection{Incorporation of Signal Penetration Depth Equation and Shannon-Hartley Theorem}
In our study, the sensor signal penetration depth equation significantly enhanced the precision with which we could understand and predict sensor signal attenuation within complex environments characterized by physical obstructions. This sensor equation was instrumental in aligning each transmitted sensor pulse relative to its predecessor, allowing for precise measurements of the differences in sensor signal arrival times. By utilizing the sensor equation

\vspace{-4mm}
\begin{equation}
S_{pd} = v \times \tau \
\end{equation}

\vspace{-2mm}
where \( S_{pd} \) denotes the signal penetration depth, \( v \) represents the velocity of the signal (often assumed to be the speed of light in free space for UWB sensor signals), and \( \tau \) is the measured time delay between the emission and reception of each pulse, we could accurately gauge how different materials and spatial configurations within the laboratory environment affected the UWB sensor signals. 

The application of this equation enabled us to refine the sensor system settings to optimize signal clarity and reception quality under varied conditions. For instance, by quantifying how much signal attenuation occurred as pulses encountered various barriers, we could make informed adjustments to the transmitter power levels and the placement of receiver sensors to counteract these effects. Furthermore, this methodological approach facilitated a predictive model of signal behavior, improving our sensor system’s ability to perform reliably in non-ideal conditions by anticipating potential areas of signal degradation and adjusting the tracking algorithms accordingly. 

Shannon–Hartley theorem was instrumental in our experimental setup, particularly for enhancing the performance of our UWB sensor tracking system in a noisy laboratory environment. The theorem was applied to quantify the maximum achievable data rate within our sensor system's inherently noisy communication channel. By calculating the channel capacity, we could optimize the bandwidth usage, thereby enhancing signal clarity and reducing transmission errors, which is crucial for the precise localization of the subject within the designated space.

\vspace{-3mm}
\begin{equation}
C = B \log_2\left(1 + \frac{S}{N}\right)
\end{equation}

\vspace{-2mm}
where \(C\) represents the channel capacity in bits per second (bps), \(B\) is the bandwidth in hertz (Hz), \(S\) is the signal power, and \(N\) is the noise power. The expression \(\log_2\left(1 + \frac{S}{N}\right)\) defines the maximum data rate achievable given the SNR, which quantifies the quality of the sensor signal relative to the background noise. This optimization was particularly effective in minimizing the impact of noise, which is prevalent in dynamic and obstacle-rich environments such as biomedical laboratories where our sensor-based experiments were conducted. By employing this theorem, we ensured that our UWB sensor tracking system operated within the best possible parameters, thereby maximizing sensor data integrity and minimizing errors in the localization of the animal. This theoretical approach not only bolstered the reliability of our sensor tracking data but also provided a deeper understanding of the practical limitations and capabilities of our UWB sensor system under various environmental conditions.

\vspace{-4mm}
\subsection{Manual Annotation of Video Data and Comparison with LoS and NLoS}
In our investigation, the process of manually annotating the video required rigorous frame-by-frame (fbf) scrutiny to ascertain significant positional changes, particularly focusing on pivotal turns made by the tracked object. Given the substantial volume of data generated by a 3-minute video recorded at a frame rate of 30 frames per second (fps), which cumulatively amounts to approximately 5,400 frames, it was imperative to strategically select moments that were critical for evaluating the sensor-based tracking technology under varied conditions. For the purpose of our analysis, we extracted 36 key reference points from the entire duration of the video. These points were judiciously chosen to represent significant behavioral alterations and movement trajectories of the object, which are essential for a comprehensive assessment of the sensor system's tracking accuracy in both LoS and NLoS scenarios. The selected frames specifically highlighted the moments when the object underwent turns or exhibited notable changes in movement, which were hypothesized to challenge the sensor-based tracking system’s accuracy.

The annotation process was facilitated using VLC media player \cite{b37}, a widely accessible and versatile video software that supports fbr analysis. This tool enabled precise location pinpointing within each selected frame, which was subsequently mapped onto a meticulously prepared coordinate system established within the experimental setup. This mapping was crucial, as it provided the ground truth data against which the UWB sensor system’s measurements were evaluated. The coordinates derived from these video annotations were utilized as the benchmark for comparing the positional data obtained from the UWB sensor system. By synchronizing the timestamps from the UWB sensor data with those of the video, we systematically analyzed the deviations in positional data under both LoS and NLoS conditions. The mean errors and their corresponding standard deviations were computed based on the discrepancies between the UWB sensor system’s outputs and the manually annotated video data.

This detailed and methodical approach to video annotation and sensor data comparison was instrumental in quantifying the error margins and offered profound insights into the operational constraints and capabilities of UWB sensor-based tracking technologies in dynamic and obstructed environments. The meticulous selection of reference points and the utilization of VLC for frame-specific analyses underscored the robustness of our methodology, providing a reliable foundation for evaluating the precision and reliability of sensor-based tracking systems in real-world applications.

\vspace{-4mm}
\section{Discussion}
\label{sec:discussion}

This study demonstrates significant advancements in the application of UWB sensor technology for high-precision tracking in controlled laboratory environments, specifically using the DecaWave DWM1001C sensor module. The experimental setup involved strategically placing UWB sensor anchors around a 24" by 48" enclosure, typical of a biomedical laboratory setting. A bio-medically tagged rat served as a dynamic subject, with its movements tracked to the nearest centimeter using an Android-based sensor tracking application. This setup mimicked real-world environments where clutter and obstructions challenge precision sensor systems. Several methodological innovations were introduced in this study. The application of the \textit{Signal Penetration Depth Equation} enabled precise synchronization of sensor signal pulses, allowing accurate measurement of time differences between unobstructed and obstructed signals. This method was crucial for modeling sensor signal strength and enhancing system reliability, particularly in obstacle-rich environments. Additionally, the \textit{Shannon-Hartley Theorem} optimized sensor channel capacity by minimizing noise interference, improving data rate, range, and reliability in sensor-based tracking. A key metric for system evaluation was the SNR quantifying in dB. SNR provided insight into the clarity and reliability of the transmitted sensor data, especially in challenging NLoS conditions. By calculating SNR, we assessed the system's resilience to environmental noise, ensuring accurate sensor localization. Empirical results confirmed the effectiveness of the proposed methodologies. Under LoS conditions (78 entries), the system achieved high precision, with an average localization error of 0.162 m (\(\sigma = 0.076\) m). In NLoS conditions, the error increased to 0.356 m (\(\sigma = 0.270\) m) due to physical obstructions. SNR was instrumental in quantifying obstruction impacts on sensor signal quality, demonstrating the system’s ability to maintain reliable sensor performance in complex environments.

These findings highlight the importance of advanced signal processing techniques, such as the \textit{Shannon-Hartley Theorem} and SNR analysis, in overcoming the limitations of cluttered environments. UWB sensor technology shows great promise for precise real-time tracking, particularly in biomedical settings where accuracy and reliability are essential.

\vspace{-4mm}
\section{Conclusion}

The potential applications of Ultra-Wideband (UWB) sensor technology in research are vast and continually evolving. Future studies are expected to extend the capabilities demonstrated here by enabling more complex behavioral analyses, particularly through the incorporation of multi-animal sensor interactions within simulated natural environments. Such advancements could include live sensor tracking of animals, such as rats, in subterranean models, providing richer datasets that closely mirror real-world ecological scenarios. Ongoing improvements in UWB sensor technology—ranging from advancements in signal processing, sensor antenna design, to the miniaturization of sensor tags and anchors—are poised to enhance its versatility across research contexts. These innovations will increase the precision of sensor localization systems, contributing to the development of comprehensive models of animal behavior and offering novel insights into animal movement and interaction.

The aim is to extend UWB sensor applications to diverse ecological environments, transforming precision sensor tracking systems and methodologies in behavioral research. This integration of sensor technology with ecological studies has the potential to reshape our understanding of animal behavior, enriching life sciences with unprecedented accuracy in sensor-based data collection and analysis.

In summary, this study demonstrates the effectiveness of UWB sensor technology for tracking and behavioral analysis in laboratory settings, while also highlighting its potential to revolutionize research across disciplines. As UWB sensor technology advances, it will likely catalyze innovations in ecological and behavioral studies, significantly enhancing data accuracy and analytical capabilities.

\vspace{-15mm}

\begin{IEEEbiography}[{\includegraphics[width=1in,height=1.25in,clip,keepaspectratio]{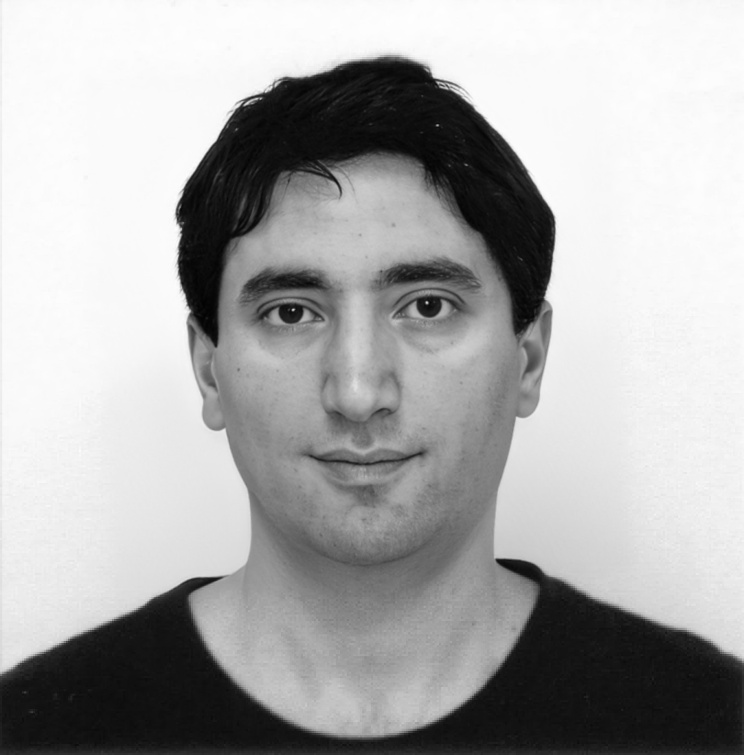}}]{Reza Sayfoori} received the B.Sc. degree from the University of South Florida (USF), Tampa, FL, USA, and the M.Sc. degree from the University of California, Irvine (UCI), Irvine, CA, USA. He is currently pursuing the Ph.D. degree in Electrical Engineering and Computer Science with the University of California (UC) at Irvine. His research interests encompass Ultra-Wideband technology for precise animal tracking, with a particular focus on its biomedical applications.
\end{IEEEbiography}

\vspace{-15mm}

\begin{IEEEbiography}[{\includegraphics[width=1in,height=1.25in,clip,keepaspectratio]{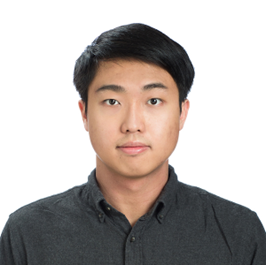}}]{Mao-Hsiang Huang} is currently pursuing the Ph.D. degree in computer engineering with the University of California (UC) at Irvine, Irvine, CA, USA. His research interests encompass the areas of machine learning and computer vision, with a particular focus on their biomedical applications.

\end{IEEEbiography}

\vspace{-15mm}

\begin{IEEEbiography}[{\includegraphics[width=1in,height=1.25in,clip,keepaspectratio]{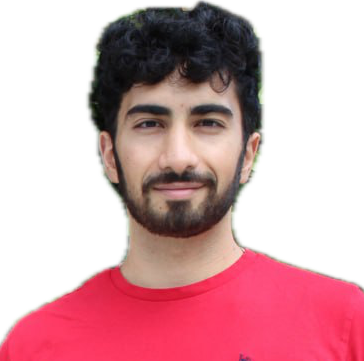}}]{Amir Naderi} \raggedright received his Ph.D. degree in computer engineering with the University of California (UC) at Irvine, Irvine, CA, USA
\end{IEEEbiography}

\vspace{-15mm}

\begin{IEEEbiography}[{\includegraphics[width=1in,height=1.25in,clip,keepaspectratio]{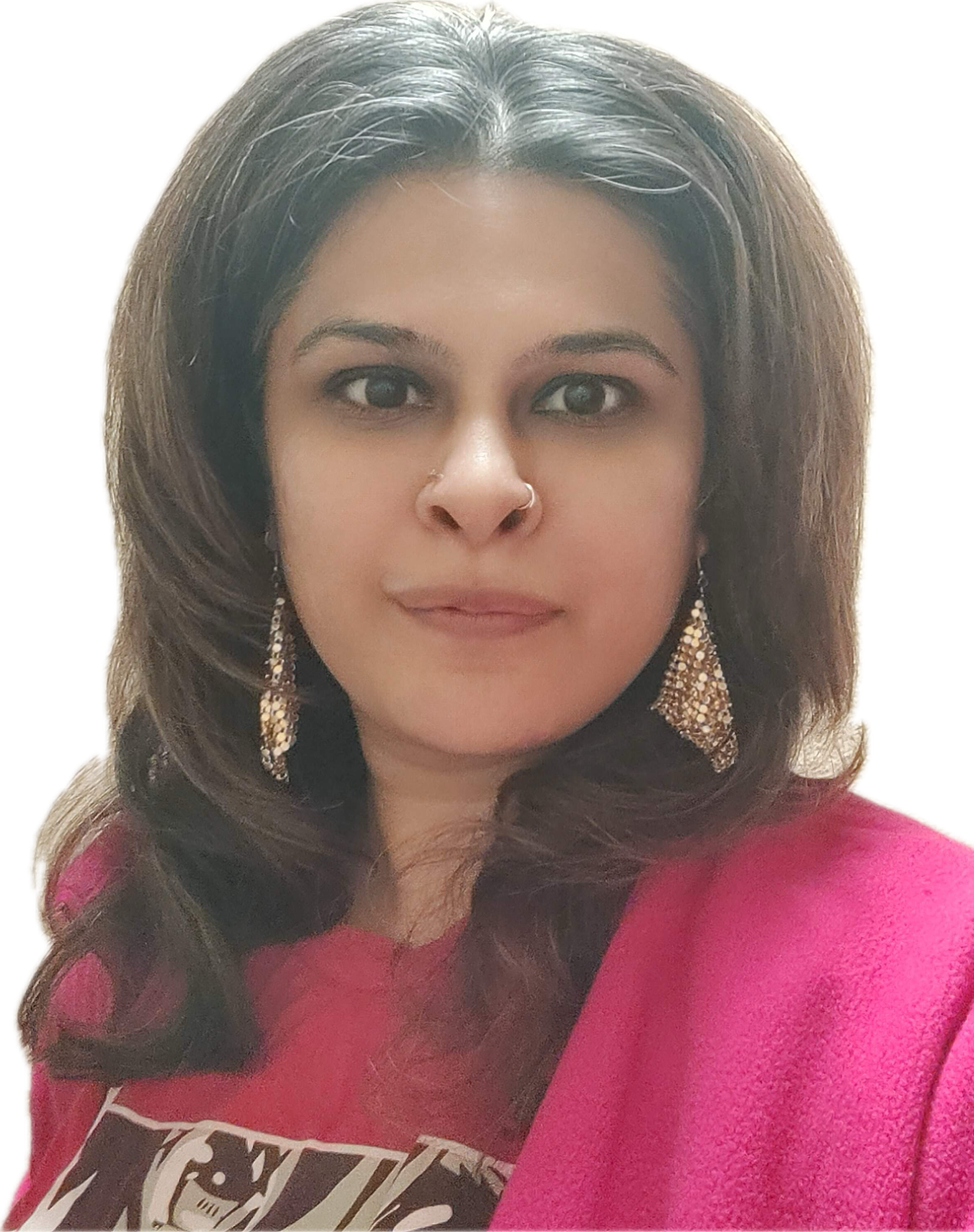}}]{Mehwish Bhatti} obtained her Ph.D. in Electrical and Electronics Engineering, University Technology PETRONAS, Malaysia; her M.Sc. in Biology and Brain Engineering, Korea Advance Institute of Science and Technology, South Korea; and her B.Sc in Computer System Engineering, G.I.K Institute of Science and Technology. She is a neuroscientist with a keen interest in understanding health and pathology of the brain. Using preclinical models of disease, she has been trying to understand the interactions of neuronal, glial and vascular interactions using neuroimaging, electrophysiology recordings, biosensors, histology, immunohistochemistry and microscopy . She is also interested in analyzing for biomarkers using computational methods, statistical analysis and machine learning.
\end{IEEEbiography}

\vspace{-15mm}
\begin{IEEEbiography}
{Ron D. Frostig}
is a Professor of Neurobiology and Behavior and Biomedical Engineering at UCI. He received his undergraduate degree in Biology, MSc. in Neurobiology, a PhD. in Neuroscience from UCLA, and a postdoc training in Neurobiology at Rockefeller University prior to his move to UCI. His main research interests include structure, function, and plasticity of neocortex, applied translational preclinical studies, and social neuroscience.
\end{IEEEbiography}

\vspace{-15mm}

\begin{IEEEbiography}[{\includegraphics[width=1in,height=1.25in,clip,keepaspectratio]{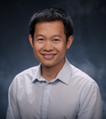}}]{Hung Cao} (Senior Member, IEEE) received the B.Sc. degree from Hanoi University of Science and Technology, Hanoi, Vietnam, in 2003 and the M.Sc. and Ph.D. degrees in electrical engineering from the UT Arlington, Arlington, TX, USA, in 2007 and 2012, respectively. He received additional training in bioengineering and medicine at the University of Southern California, Los Angeles, CA, USA, from 2012 to 2013; and the University of California at Los Angeles, Los Angeles, from 2013 to 2014. From 2014 to 2015, he worked for ETS, Montreal, QC, Canada, as a Research Faculty. From 2015 to 2018, he worked as an Assistant Professor of electrical and biomedical engineering at the University of Washington Bothell, Bothell, WA, USA. He joined the University of California (UC) at Irvine, Irvine, CA, USA, in 2018, where he is currently an Associate Professor of electrical engineering, biomedical engineering, and computer science. Dr. Cao was a recipient of the National Science Foundation (NSF) CAREER Award in 2017.
\end{IEEEbiography}
\vspace{-15mm}
\end{document}